\documentclass[useAMS,usenatbib]{mn2e}
\usepackage{psfig}
\usepackage{epsfig}
\usepackage{references}
\bibpunct{(}{)}{;}{a}{}{,}

\def\dn{$\Delta\nu$}

\def\qseg{Q$_{\rm seg}$}
\begin{document}
\title {An abrupt drop in the coherence of the lower
   kilohertz QPO in 4U 1636-536}
    \author[Barret, Olive \& Miller]{Didier Barret$^{1}$\thanks{E-mail:
    Didier.Barret@cesr.fr}, Jean-Francois
    Olive$^{1}$ \& M. Coleman Miller$^{2}$ \\
$^1$Centre d'Etude Spatiale des Rayonnements, CNRS/UPS, 9 Avenue du
Colonel Roche, 31028 Toulouse Cedex 04, France\\
$^2$Department of Astronomy, University of Maryland, College Park, MD
20742-2421, United States}
\date{Accepted 2005 XX. Received 2005 XX; in original 2005 XX}

\pagerange{\pageref{firstpage}--\pageref{lastpage}} \pubyear{2005}
\maketitle

\label{firstpage}

\begin{abstract}
Using archival data from the Rossi X-ray Timing Explorer (RXTE), we
study in a systematic way the variation of the quality factor (Q=$\nu/\Delta\nu$,
$\Delta\nu$, FWHM) and amplitude
of the lower and upper kHz quasi-periodic oscillations (QPO) in the
low-mass X-ray binary 4U1636-536, over a frequency range from $\sim 550$ Hz to
$\sim 1200$ Hz.  When represented in a quality
factor versus frequency diagram, the upper and lower QPOs follow two
different tracks, suggesting that they are distinct phenomena, although
not completely independent because the frequency difference of the two
QPOs, when detected simultaneously, remains within $\sim 60$ Hz of half the neutron
star spin frequency (at $\nu_{\rm spin}=581$ Hz).  The quality factor  of the lower kHz QPO increases with frequency up to a
maximum of $Q\approx 200$ at $\nu_{\rm lower}\approx 850$ Hz, then drops
precipitously to $Q\approx 50$ at the highest detected frequencies
$\nu_{\rm lower}\approx 920$~Hz. A ceiling of the lower QPO frequencies at 920 Hz is also clearly seen in a frequency versus count rate diagram.  At the same time, the quality factor
of the upper kHz QPO increases steadily from $\sim 5$ to $Q\sim 15$ all the way to
$\nu_{\rm upper}\approx 1150$ Hz, which is the highest detectable QPO frequency.  The rms amplitudes of both the upper and lower kHz QPOs decrease steadily towards higher frequencies. 

The quality factor provides a measure of the coherence of the underlying oscillator. For exponentially damped sinusoidal shots, the highest Q observed corresponds to an oscillator coherence time of $1/\pi\Delta\nu\sim 0.1$ seconds.  All existing QPO models face challenges in explaining such a long coherence time and the significantly different behaviours of
the quality factors of the upper and lower QPOs reported here. 
It is therefore difficult to be certain of the implications of the
abrupt change in the lower QPO at $\sim 850$~Hz. We discuss various possible causes, including that the drop
   in coherence is ultimately caused by effects related to the
   innermost stable circular orbit.
\end{abstract}

 
\begin{figure*}
\includegraphics[width=0.75\textwidth]{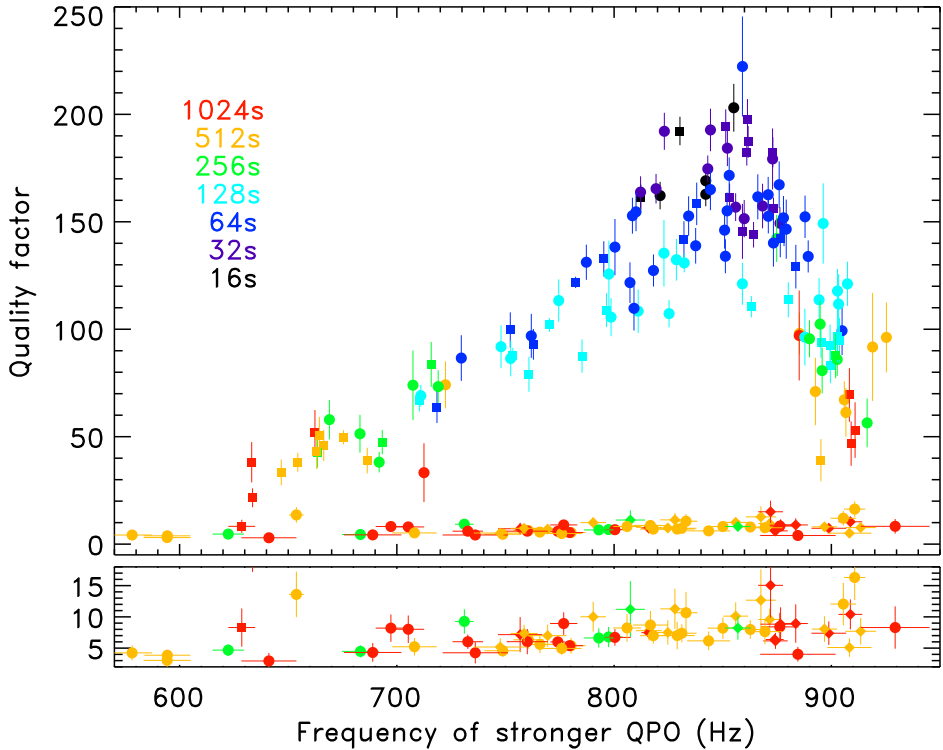}
\caption[]{Top panel: Quality factor versus frequency of the stronger 201 QPOs detected from 4U1636-536. There are clearly two classes of QPOs
showing up in that figure; one for which Q steadily increases between
600 and 850 Hz and then decreases afterwards (this branch contains
all identified lower QPOs, filled squares), and another class of QPOs which
includes those detected with the longest integration times, and which
seems to show a positive correlation between Q and frequency between 600
Hz and 900 Hz (it includes all identified upper QPOs, filled diamonds).
Segments for which only one QPO was detected are shown with filled circles. They are mostly found on the
lower QPO branch. Different colors correspond to different integration
times for the PDS, ranging from 16 to 1024 seconds. Bottom panel: A blow-up of the Y axis suggesting that for the upper QPO, a positive correlation exists between Q and frequency over the frequency range  $\sim 600$ to 930 Hz (the Spearman correlation coefficient is 0.35 corresponding to a null-hypothesis probability of $3.4~10^{-7}$, see also Figure 4). }
\label{barret_f1}
\end{figure*}

\section{Introduction}
Following the discovery with RXTE in 1996 of kHz QPOs in neutron star
low-mass X-ray binaries (LXMBs), it was immediately realised that
they offer unique probes of the strongly curved spacetime around
compact stars, possibly providing ways of testing some of the most fundamental predictions of General Relativity (GR) in the strong field regime (e.g., van der Klis 2005).  One of the key predictions
   of strong gravity GR is that there exists a region around sufficiently
   compact stars within which no stable circular orbital motion is possible.
   In a Schwarzschild geometry, the radius of the innermost stable circular
   orbit (ISCO) is $6GM/c^2$ (e.g., Bardeen et al. 1972; see also 
   Klu\'zniak \& Wagoner 1985), which is greater than the radii of
   neutron stars constructed with most modern high-density equations 
   of state (see Akmal, Pandharipande, \& Ravenhall 1998).  Prior to the launch of
RXTE, it was suggested that the ISCO might induce a frequency cutoff
in the power density spectrum (Klu\'zniak et al. 1990).  After the
discovery of kHz QPOs, it was proposed that signatures of the ISCO
could include a frequency saturation of kHz QPOs with increasing mass
accretion rate, or a drop in the quality factor and rms amplitude of
kHz QPOs as they approach a limiting frequency (Miller et al. 1998).

 An apparent frequency saturation with increasing countrate was in
   fact claimed for the neutron star LMXB 4U~1820--30 (Smale, Zhang,
   \& White 1997; Zhang et al. 1998).  This was disputed due to 
   concern that the countrate might not be an accurate measure of
   mass accretion rate, and it was proposed that the quantity $S_a$
   (a measure of the location of an atoll source on an X-ray color-color
   diagram) would be more representative (M\'endez et al. 1999; M\'endez
   2002).  Additional analyses of data integrated over RXTE orbits
   show frequency saturation against many other measures
   of mass accretion rate, including $S_a$ (e.g., Kaaret et al. 1999;
   Bloser et al. 2000).  However, short-term variations of the frequency
   are important, and there is not a one-to-one correlation between $S_a$ and the QPO frequency (e.g. di Salvo et al. 2003), suggesting that $S_a$ might not be a precise indicator of mass
 accretion rate after all.
   Therefore, although the ISCO interpretation in 4U~1820--30 remains
   a possibility, its status is uncertain.
   
The large RXTE archive now makes possible searches for systematic
correlations between frequency, quality factor, and amplitude in
unbiased data sets.  For example, our analysis of the neutron star
LMXB 4U 1608--52 revealed a positive correlation between the frequency
$\nu_{\rm lower}$ of the lower frequency kHz QPO and its quality factor,
up to a maximum of about $Q=200$, beyond which the two quantities were
anticorrelated (Barret et al. 2005).  Such a result places stringent
constraints on models, especially those relating QPOs to orbiting clumps.

Motivated by this result, we investigate the properties of QPOs in the
similar system 4U 1636--536.  This system is ideal for several reasons:
its kHz QPOs are strong, their frequencies vary over a wide range, and 4U
1636--536 has been observed extensively with RXTE. Its
kHz QPO properties have been studied as a function of count rate,
luminosity, energy spectral parameters, and other quantities (e.g., Zhang
et al. 1996a,b; Wijnands et al. 1997; Jonker, M\'endez, \& van der Klis
2000, 2002; Di Salvo, M\'endez, \& van der Klis 2003; Misra \& Shanthi
2004).  

Here we focus on the quality factor and amplitude of the kHz
QPOs, which we analyse systematically by putting together all the
available observations of the RXTE archive to date.
We discuss our data reduction procedure in \S~2.
We discuss the possible implications of our results
in \S~3. 
   
\section{Data reduction procedure}
We have retrieved all the Proportional Counter Array science event files
with time resolution better than or equal to 250 micro-seconds. Only data
files with exposure times larger than 600 seconds are considered here.
The data set covers the period from  April 1996 to September 2004. No
filtering on the raw data is performed, which means that all photons are
used in the analysis, only time intervals containing X-ray bursts are removed.
We have computed Leahy normalised Fourier power density spectra (PDS)
between 1 and 2048 Hz over 8 s intervals (with a 1 Hz resolution).

\subsection{QPO identification}

A PDS which is the average of all the 8s PDS over a continuous data segment is first
computed; the typical length of the segment is $\sim 4500$ seconds,
i.e. comparable to the orbital period of the RXTE spacecraft. The
segment-averaged PDS is then searched for a QPO using a scanning
technique which looks for peak excesses above the Poisson counting
noise level (see Boirin et al. 2000). Because at the lowest frequency,
typically $\sim 400$ Hz, QPOs become broad features, merging with broad
noise components (e.g. Olive et al. 2003), they
are difficult to detect with an automatic procedure. For this reason,
the QPO search is restricted to frequencies above 500 Hz. The detection threshold for
the QPO is $8\sigma$ in a given segment. Of the 631 event files analyzed,
201 matched our selection criteria.

Having located the strongest QPO peak in each segment-averaged PDS, we
now wish to get an estimate of the quality factor of the QPO, reducing
as much as possible the contribution of the long term frequency
drifts. We therefore follow the QPO on the shortest possible
timescales permitted by the data statistics. We divide each continuous
data segment into N intervals, and search for a QPO within a 50 Hz
window around the frequency of the QPO in the segment-averaged PDS. The possible interval durations considered are 16 s, 32 s, 64 s, 128 s,
256 s, 512 s, and 1024 s. The shortest usable integration time is then
estimated such that the QPO is detected (above $4\sigma$) in at least
80\% of the  N intervals (a linear interpolation is used to estimate
the QPO frequency in the remaining intervals).  The N PDS are then (frequency)
shifted to the mean QPO frequency over the segment and averaged (M{\'
e}ndez et al. 1998b). The segment-averaged QPO profile is then fitted
over a 100 Hz frequency range (50 Hz on each side of the QPO peak)
with a Lorentzian having three parameters (amplitude, frequency and
\dn)  plus a constant representing the counting-statistics noise
level, following the method described in Barret et al. (2005). The
uncertainties in the fitted parameters are computed such that $\chi^2=\chi^2_{\rm min} +1$ for variation of one single parameter.  

The quality factor of the strongest QPOs (\qseg), so obtained, is represented in  Figure 1. This figure shows two distinct branches. For the main branch \qseg~increases with frequency up to a maximum ($\sim 850$ Hz) beyond which a sharp decrease is observed. A bottom track is seen at the bottom of the plot.  It includes broader QPOs, detected only on the longest integration times. There is a trend for \qseg~of these QPOs to increase with frequency along the track. 

To determine whether the strongest QPO detected is the lower or the upper kHz QPO, the shifted and averaged PDS for each segment is searched for a second QPO using the same scanning technique between 500 and 1300 Hz. A second QPO is detected if its significance above the counting noise level is greater than $4\sigma$ (following the method described in Boirin et al. 2000). With this procedure, we directly detect a weaker upper kHz QPO in 53 segments: all segments marked with filled squares in Figure 1 are located on the upper branch. We also detect a weaker lower kHz QPO in 24 segments: all of these segments marked with filled diamonds are located on the bottom branch of Figure 1. From this we conclude that the upper curvy branch (\qseg$>15$) is the lower QPO branch, whereas the bottom track  (\qseg$<15$) is the upper QPO branch. As can be seen from Figure 1, many segments including a single QPO (filled circles) (i.e. in which the second one is not detected above $4\sigma$) are found on the lower QPO branch, hence contain lower QPOs. This demonstrates that our method, applied to a large data set, provides a way to identify kHz QPOs, based on the frequency dependency of their Q values. Alternative methods do exist, as for instance the one based on frequency correlations between kHz QPOs and features at lower frequencies $<$ 100 Hz (van Straaten et al. 2003).

 Note that the upper QPO is detected as the strongest QPO when its frequency is between $\sim 600$ and 930 Hz. This means that there are no segments of data where the upper QPO has a frequency larger than 930 Hz and a significance larger than $8 \sigma$. Obviously, this does not mean that the upper QPO does not reach frequencies higher than 930 Hz, as we will see below. Over a limited frequency range (770 Hz to 920 Hz), and for a limited data set, the same trends for the quality factor of both QPOs were present in the 4U1636-536 data analysed in \cite{disalvo03}.

In the above analysis, the integration time used in estimating \qseg\ is minimized and therefore is not the same for all the PDS. It ranges from 16 to 1024 seconds, leading to a
potential bias, because weaker or broader QPOs will require longer
integration times to be detected, and hence will suffer from a larger
broadening due to a larger frequency drift within the interval. As
shown in Figure 1, the highest \qseg\ are indeed detected on the shortest
time scales. We have verified for both the upper and lower QPOs that the same trend is observed for their quality factors when the PDS integration time is not optimized, but instead is the same for all segments, e.g. 1024 seconds. The upper and lower branches are still clearly distinct; the main difference being that the upper one (which contains segments for which the PDS integration time could be reduced to 16 seconds in some cases) is shifted downwards to smaller \qseg\ (with a maximum around 100). This is expected as the QPO width includes now a larger contribution from the long term frequency drifts, leading to smaller \qseg.

More than 7900 individual PDS are used in the present analysis, amounting for about 840 kilo-seconds of QPOs. The detected QPO frequencies in the individual PDS  are plotted against total source count rates in Figure 2 for the lower QPO as identified above (those corresponding to \qseg\ $>$ 15). This figure shows the so-called parallel track phenomenon (M\'endez et al. 1999, van der Klis 2001). The tracks flatten at high count rates. Note that there is a clear ceiling in frequency (around 920 Hz) above which no QPOs are detected, whatever the source count rate is.

\begin{figure} 
\includegraphics[width=0.475\textwidth]{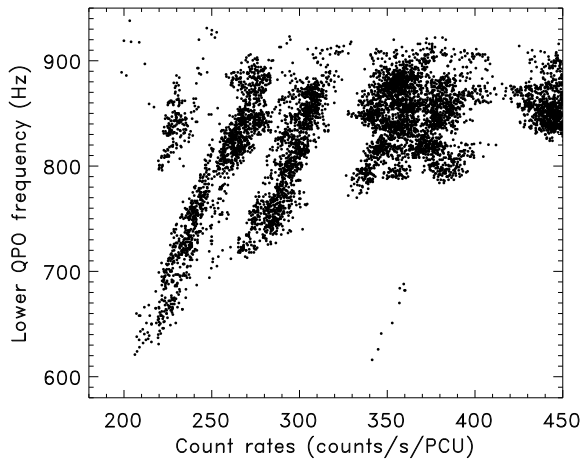}
\caption[]{Lower QPO frequencies versus count rates (normalized by the number of PCA units operating). Note the clean limitation in frequency around 920 Hz seen for the lower QPOs.}
\label{barret_f2}
\end{figure}
\subsection{Q and amplitude versus frequency}
In order to get a better estimate of the Q versus frequency dependency, we now divide the individual PDS in two classes, those contributing to  \qseg~larger than 15 (the lower QPOs) and those associated with \qseg~smaller than 15 (the upper QPOs).  The lower QPO spans a frequency range from 620 Hz to 920 Hz. We divide this range with steps of 15 Hz. In each frequency bin (e.g. 620-635 Hz), all the individual PDS whose strong QPO frequency fall in the bin are shifted-and-added to a mean QPO frequency measured within the bin. The lower QPO is then fitted, and its quality factor and rms amplitude estimated from the fit. An upper QPO is always present. However, it is broad ($\ga$ 70 Hz) and therefore difficult to fit. We could obtain reliable fit parameters for 18 segments. We repeat the analysis for the PDS associated with \qseg~less than 15, but with frequency bins of 30 Hz (as we have less individual PDS involved). A significant fit for the lower QPO is obtained on only 2 occasions. 

Figure 3 and 4 show the quality factor and rms amplitude of the lower and upper QPOs as estimated with our analysis. The patterns seen for Q in Figure 1 are evident, but this time with much less scatter.  The most striking features of these two figures are the following: the quality factor of the lower QPO shows a smooth increases with frequency, reaching a maximum at 850 Hz, and decreasing sharply afterwards. On the other hand, the quality factor of the upper QPO smoothly increases up to 850 Hz, with no evidence for a drop afterwards (note however that only the narrower signals have been fitted and the error bars are relatively large). Finally, the rms amplitude of both QPOs decreases steadily at the highest frequencies. The upper QPO is detected close to the rms detection threshold of our analysis: typically $\sim 2.0$\% rms ($3\sigma$) for a typical width of 50-100 Hz. Therefore its non-detection at frequencies higher than $\sim 1150$ Hz could be due to a lack of sensitivity\footnote{Note however that if the lower QPO frequency is limited to $\sim 920$ Hz with the frequency difference between the upper and lower QPOs remaining less than half the spin frequency, as suggested in Figure 5, the upper QPO frequency is not expected to reach values higher than $\sim 1150-1200$ Hz.}. On the other hand, for the lower and narrower QPO, our detection threshold is closer to $\sim 1.0$\% rms. This value is less than one fourth of the rms measured for its highest frequency at $\sim 920$ Hz, thus providing further support to the idea that the lower QPO truly disappears.
\begin{figure} 
\includegraphics[width=0.475\textwidth]{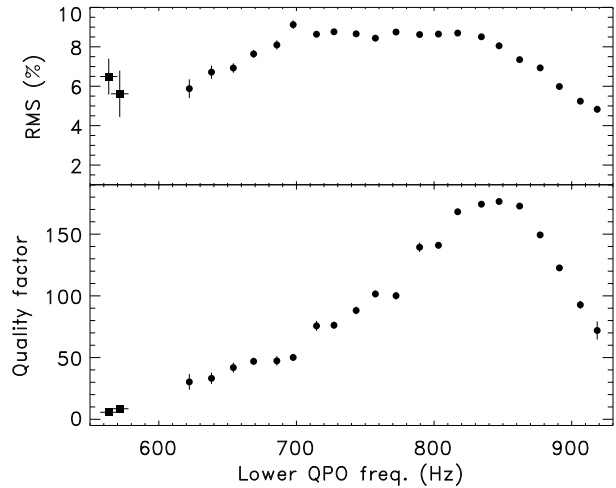}
\caption[]{rms amplitude and quality factor of the lower kHz QPO of 4U1636-536, as measured every 15 Hz. Points marked with filled squares have been obtained from fitting the lower QPO in segments where the upper QPO was the strongest.}
\label{barret_3}
\end{figure}

\begin{figure} 
\includegraphics[width=0.475\textwidth]{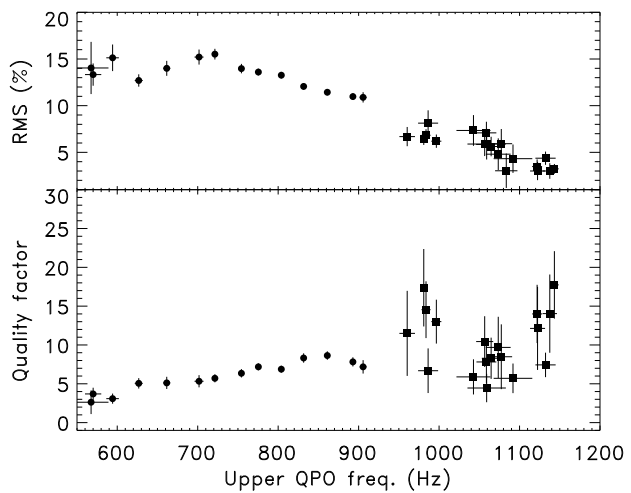}
\caption[]{rms amplitude and quality factor of the upper kHz QPO of 4U1636-536 as derived every 30 Hz. Points marked with filled squares have been obtained from fitting the upper QPO in segments where the lower QPO was the strongest.}
\label{barret_3}
\end{figure}

\subsection{Relation with spin}
From  X-ray burst oscillations, the spin frequency ($\nu_{\rm spin}$) of the
neutron star is thought to be 581 Hz (Strohmayer \& Markwardt 2002). As shown in Figure 5, when we detect both QPOs in the above analysis, the frequency difference remains within $\sim 60$ Hz of half the spin frequency, with a mean value around 300 Hz (our results are fully consistent with the one presented in Jonker et al. (2002)). It is interesting to note that the peak of coherence is reached when the frequency difference is the closest to half the spin frequency.
\section{Discussion}
\begin{figure} 

\includegraphics[width=0.475\textwidth]{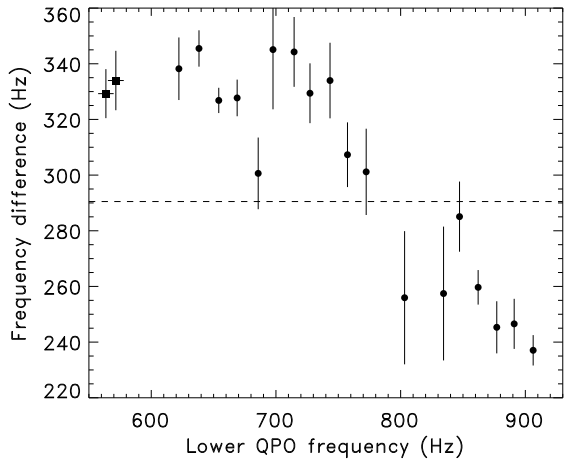}
\caption[]{Frequency difference between the upper and lower QPOs (filled squares correspond to points when the upper is the strongest QPO of the two, filled circles are those for which the lower is the strongest QPO of the two). Half the neutron star frequency is indicated with a dashed line at 290 Hz. The frequency difference remains within $\sim 60$ Hz of half the spin frequency, suggesting that the spin plays a role in setting up the frequencies.}
\label{barret_3}
\end{figure}

We have studied systematically the quality factor and rms amplitude of
the kHz QPOs detected from 4U~1636--536 with the Rossi X-ray Timing
Explorer.  Using a shift-and-add technique to minimize the contribution
of the long term frequency drift to the measured width of the QPO, we
have shown that the lower and upper kHz QPOs follow two different tracks
in a diagram of quality factor versus frequency. 

The most striking result reported here is the frequency evolution of
the coherence of the lower kHz QPOs.  The coherence of the lower QPO shows a
clear increase from 600 Hz to 850 Hz, then a sharp decrease in
coherence to the maximum observed frequency of about 920 Hz (see Figure
3).  The same behaviour was suggested from the 4U 1608--522 data
(Barret et al. 2005).  The rms amplitude remains
roughly constant between 600 Hz and 820 Hz and then decreases
smoothly.  A similar dependence of rms amplitude on frequency has been
reported for other systems (M\'endez, van der Klis, \& Ford 2001). The
quality factor of the lower kHz QPO reaches Q=222$\pm$24 at 860 Hz on
12 August 2000. This value, as well as the frequency at which it
occurs, is remarkably similar to the one observed previously from 4U
1608--522 (Berger et al. 1996; Barret et al. 2005).  For exponentially
damped sinusoidal shots, this high Q value corresponds to an oscillator
coherence time of $1/\pi\Delta\nu\sim 0.1$ seconds, or about 70 cycles.  As
discussed in Barret et al. (2005), this long coherence time, as well as the
different behaviours of the lower and upper QPOs (see Fig. 1, 3, 4), pose severe
constraints on {\it all} existing QPO models. These findings are particularly constraining on models
involving orbiting clumps; the reason being that clumps, left alone in
the disk, will be sheared by differential rotation in just a few
cycles. For the clumps to survive against tidal shear would require that some
local forces, unidentified so far, can keep them together.

Our analysis of 4U 1636--536 and 4U 1608--52, combined with previous
results on kHz QPOs, produces an apparent paradox.  The different
quality factor dependences of the two QPOs might seem to
imply that the mechanisms for generating the two QPOs are
decoupled from each other.  However, it has been shown in many systems
that the separation frequency between the lower and upper kHz QPOs is
close to the stellar spin frequency or half the spin frequency (see,
e.g., Wijnands et al. 2003 for a discussion related to the
accretion-powered millisecond pulsar SAX J1808--3658). For 4U~1636--536,
the frequency difference we measured between the two QPOs is also
close to half the neutron star spin frequency (see Figure 5). This would seem to
suggest that the two kHz QPOs are linked to each other,
and that the spin plays a role in producing at least one of the QPOs. 
Resolution of these apparently discrepant observations will be an
important task for QPO models.

Despite the uncertainties about the origin of the QPOs, some
relatively model-independent conclusions can be drawn.  Let us
first consider what governs the frequencies and coherences of the
QPOs.  There are three regions in the system that can produce
emission: the cool disk, the hot corona above the disk, and the
boundary layer of accretion onto the star (e.g. Barret et al. 2000). In fact, the presence of
QPOs in the high energy emission E$>20$ keV (van der Klis 2005) requires that at
least some emission come from other than the $kT\sim 1-2$~keV
disk.  However, this does {\it not} mean that the observed
frequency {\it originates} in the corona or boundary layer.  For
example, oscillations in the disk could modify the temperature or
optical depth of the corona, or the accretion rate onto the star,
leading to modulation of the high energy emission.  Indeed, there
are no easily conceivable mechanisms by which highly coherent
oscillations could be generated primarily in the corona, let alone
in the turbulent boundary layer.  In contrast, the disk offers many
possibilities for characterstic frequencies, from orbital or
epicyclic frequencies to oscillation modes.  We therefore consider
it highly likely that the frequencies of the QPOs are determined
in the disk, even if the emission is produced elsewhere.

Now consider the implications of a sharp change in QPO behaviour
that occurs reproducibly at a given frequency.  Such a change implies
that there is something special about the disk when that frequency
is reached.  If this change occurred only once, or just at a single
countrate, it could be that for complicated reasons the disk undergoes
a change of state.  In the case of 4U~1636--536, however, the limiting
frequency and associated coherence drop are robust against countrate
and other factors.  This argues in favor of a more stable special
frequency.  Given that in the disk, frequencies are strongly linked
to radii, we must therefore look for special radii in the system at
which abrupt changes are plausible.

There are three candidates for such a radius: the radius of the magnetosphere,
the radius of the star, and the radius of the ISCO.  In modern equations
of state for dense matter (e.g., Akmal, Pandharipande, \& Ravenhall 1998),
the star is contained well inside the ISCO, especially for stars rotating at rates similar to 4U1636-536.  This argues against the
stellar surface as a candidate for sharp frequency changes, simply
because crossing the ISCO will surely produce dramatic effects (indeed,
highly coherent oscillations should be impossible inside the ISCO), and
this crossing would occur at a lower frequency than that of an orbit
at the stellar surface.  The options are therefore the magnetosphere and
the ISCO.  It is difficult to rule out a magnetospheric explanation,
because very close to the star the magnetic field could have a
complicated geometry with effects that are difficult to anticipate.
We do note that the coherence drop is extremely abrupt (a factor of
10 in quality factor over a 8\% change in frequency), which might not
be expected from a magnetosphere unless the magnetic field is 
extremely stiff.  Nonetheless, this is a possibility.

The remaining possibility is that the ISCO drives the sudden drop in
coherence.  The sudden drop in coherence at a reproducible limiting
frequency is one of the signatures expected for the ISCO (see, e.g.,
Klu\'zniak et al. 1985, Miller et al. 1998).  The frequency at which
this occurs is also reasonable. The behaviour we see is therefore
qualitatively and quantitatively consistent with expected effects of
the ISCO. However, the complexity of the source behaviour and the lack of evidence for abrupt changes in the upper QPO as well as the lower QPO mean that it is
difficult to draw definite conclusions as to the origin of the drop in quality factor.  With those caveats, we now discuss the implication for the neutron star mass, if the observed drop is related to the ISCO.

Within the ISCO interpretation, but independent of any detailed
models, we can estimate the mass of the star. Our observations
suggest that the frequency of the lower QPO is bounded at $\sim$900~Hz,
possibly implying that the frequency of the upper QPO is also
limited to $900~{\rm Hz}+\nu_{\rm spin}/2$, or $\sim 1200$ Hz (the
highest upper QPO frequency observed is $\sim$1150 Hz).  For
a rapidly rotating star such as 4U~1636--536, which is likely to
have a dimensionless angular momentum of $j\equiv cJ/GM^2\approx
0.1-0.2$, the mass implied by an ISCO frequency of 1200~Hz is
$\approx 1.9-2.1\,M_\odot$ (see Miller et al. 1998 for the
applicable formulae).  Such a mass, although high compared to most
estimated neutron star masses, is consistent with the most current
nucleonic equations of state with plausible three-body repulsion
(see Akmal, Pandharipande, \& Ravenhall 1998), and could have
resulted from Eddington-level accretion for a few tens of millions
of years onto a neutron star with initial mass  $1.4\,M_\odot$.

\section{Conclusions}

By analysing in a homogeneous way all the available archival 
observations, we have followed the quality factor and rms amplitude of
the two kHz QPOs detected in 4U 1636--536.  One striking feature of
our findings is that, towards the highest frequency, we observe a
behaviour which may be consistent with a signal originating close to the ISCO:  a
reproducible and abrupt drop in the coherence of the lower kHz QPO.
This result was obtained because we were able, for the first time, to
analyse a very large data set homogeneously with no a priori data
selection.  The potential importance of these results demands more
accumulation of evidence, but if similar patterns are seen in other
systems then the rapid timing capabilities of RXTE may indeed have
allowed the discovery of a long-anticipated effect of strong gravity.

\section*{Acknowledgments}
We are very grateful to Gerry Skinner,  Jean-Pierre Lasota
and Wlodek Klu\'zniak for very helpful and stimulating discussions, during
the preparation of this paper. We are thankful to Mariano M\'endez for his very valuable comments on the results presented, and for sharing his expertise on QPO studies with us. We thank an anonymous referee for suggestions which helped to strengthen the results presented in this paper. MCM was supported in part by NSF grant AST
0098436 at Maryland.  MCM also thanks the Theoretical Institute for
Advanced Research in Astrophysics (Hsinchu, Taiwan) for hospitality
during part of this work.  This research has made use of data obtained
from the High Energy Astrophysics Science Archive Research Center
(HEASARC), provided by NASA's Goddard Space Flight Center.

\end{document}